\documentclass{article}
\usepackage{ijcai18}

\usepackage{times}
\usepackage{todonotes}
\usepackage{graphicx,changepage}
\usepackage{xcolor}
\usepackage{color}
\usepackage{float}
\usepackage{soul}
\usepackage{tikz}
\usepackage[utf8]{inputenc}
\usepackage[small]{caption}
\usepackage{tikz}
\usepackage{booktabs}

\def\checkmark{\tikz\fill[scale=0.4](0,.35) -- (.25,0) -- (1,.7) -- (.25,.15) -- cycle;}

\title{Closing the AI Knowledge Gap}

\author{
Ziv Epstein\thanks{Equal contribution.}, 
Blakeley H. Payne\footnotemark[1], 
Judy Hanwen Shen, 
Abhimanyu Dubey, \\
\textbf{Bjarke Felbo}, 
\textbf{Matthew Groh}, 
\textbf{Nick Obradovich}, 
\textbf{Manuel Cebrian}, 
\textbf{Iyad Rahwan}
\\Media Lab, Massachusetts Institute of Technology, Cambridge, MA, USA\\
Correspondence: \{cebrian, irahwan\}@mit.edu
}

\begin{document}

\maketitle

\begin{abstract}
AI researchers employ not only the scientific method, but also methodology from mathematics and engineering. However, the use of the scientific method -- specifically hypothesis testing -- in AI is typically conducted in service of engineering objectives. Growing interest in topics such as fairness and algorithmic bias show that engineering-focused questions only comprise a subset of the important questions about AI systems. This results in the \emph{AI Knowledge Gap}: the number of unique AI systems grows faster than the number of studies that characterize these systems' behavior.
%
%
%
To close this gap, we argue that the study of AI could benefit from the greater inclusion of researchers who are well positioned to formulate and test hypotheses about the behavior of AI systems. We examine the barriers preventing social and behavioral scientists from conducting such studies. Our diagnosis suggests that accelerating the scientific study of AI systems requires new incentives for academia and industry, mediated by new tools and institutions. To address these needs, we propose a two-sided marketplace called TuringBox. On one side, \emph{AI contributors} upload existing and novel algorithms to be studied scientifically by others. On the other side, \emph{AI examiners} develop and post machine intelligence tasks designed to evaluate and characterize algorithmic behavior. We discuss this market’s potential to democratize the scientific study of AI behavior, and thus narrow the AI Knowledge Gap.

\end{abstract}

\section{The Many Facets of AI Research}

Although AI is a sub-discipline of computer science, AI researchers do not exclusively use the scientific method in their work. For example, the methods used by early AI researchers often drew from logic, a subfield of mathematics, and are distinct from the scientific method we think of today. Indeed AI has adopted many techniques and approaches over time. In this section, we distinguish and explore the history of these methodologies with a particular emphasis on characterizing the evolving science of AI.
\subsection{AI as Math}

As early as the seventeenth century, the notion that intelligence could be equated to symbolic information processing was formalized. In his 1677 \emph{Preface to the General Science} Leibniz wrote: ``It is obvious that if we could find characters or signs suited for expressing all our thoughts as clearly and as exactly as arithmetic expresses numbers or geometry expresses lines, we could do in all matters insofar as they are subject to reasoning all that we can do in arithmetic and geometry. For all investigations which depend on reasoning would be carried out by transposing these characters and by a species of calculus.'' \cite{leibniz}

In the early 20th century, Leibniz's ideas influenced a number of mathematicians. Logician and mathematician David Hilbert posed the famous question: Can all of mathematical reasoning be formalized \cite{hilbert1928grundlagen}? This question spurred many others to explore the limits of computation and logic, an enterprise that culminated in G\"{o}del's incompleteness theorems, which revealed fundamental limits of formal reasoning.

But these discoveries of formal limitations around computability and formal reasoning did not stop scholars from pursuing the foundations of mechanized intelligence. AI, as its known today, began with Alan Turing's seminal work ``Computing Machinery and Intelligence'' in which Turing discussed the idea of creating machines that can think \cite{turing1950computing}. Turing acknowledged the vagueness of the terms `machine' and `thought,' and operationalized the question with the Universal Turing Machine, which could perform arbitrary symbolic computation. The idea, again, was that once mathematics is mechanized, all manner of reasoning subsequently follows.

Thus, the early builders of AI systems were mainly mathematicians, devising mechanistic procedures--often called proof theories--for all manner of reasoning. In 1955, Herbert Simon and Allen Newell's \emph{Logic Theorist} proved 38 theorems in the \textit{Principia Mathematica}  \cite{newell1959report}. This led Simon to claim that they had ``solved the mind-body problem.'' He argued that with a sufficiently powerful version of the Logic Theorist, we could automate mathematical reasoning, which in turn would enable the automation of all reasoning.

In subsequent decades, theoretical developments in symbolic reasoning continued. This manifested in numerous  mathematical investigations, from analyzing the computational complexity of various symbolic reasoning problems, to providing precise mathematical semantics of logic programming languages \cite{van1976semantics}, to new forms of computationally efficient symbolic reasoning on constrained languages \cite{mccarthy1981circumscription,jaffar1994constraint,kakas1992abductive}. In tandem, significant theoretical developments took place in machine learning, grounded in statistical learning theory. These efforts resulted in rigorous foundations for reasoning about facts \cite{pearl1986fusion}, learning patterns \cite{vapnik1999overview} and taking actions and planning in the presence of uncertainty~\cite{sutton1998introduction} .

\subsection{AI as Engineering}
Since the 1970s, in parallel to these developments in AI theory, a growing cadre of AI engineers started taking shape. On one hand, some mathematically-minded computer scientists produced demonstrations of the capabilities of particular symbolic reasoning techniques. For example, one can show how an AI agent performs planning by solving logical satisfiability problems \cite{kautz1992planning} or performs various commonsense reasoning tasks using circumscription-based logical reasoning  \cite{mccarthy1986applications}. 

On the other hand, AI engineers embarked on formalizing all manner of domain-specific facts in symbols and rules that can be used in operational expert systems \cite{jackson1998introduction}. There were ambitious -- but ultimately unsuccessful --  attempts to build expert systems manually with encyclopedic knowledge, capable of answering any question \cite{lenat1985cyc}. The difficulty of curating such knowledge bases made salient that building general AI systems was not only a challenge from a computational perspective, but also required an infeasible knowledge engineering effort.

Despite two `AI Winters,' the engineering methodologies employed by builders of AI systems became increasingly sophisticated, leveraging contemporaneous increases in computation capacities and data streams. This sophistication took the form of i) better developed methodologies, and ii) more standardized and precise benchmarks.
%
%

First, AI engineers developed standardized knowledge engineering methodologies for AI systems, such as KADS \cite{wielinga1992kads}. This development also benefited from the overall maturity in the broader field of software engineering, enabling standardized Application Programming Interfaces, team management, code documentation and sharing, and so on. 

Second, AI engineers have developed increasingly sophisticated benchmark problems to compare their systems. These benchmarks have taken three forms: as standardized tasks, datasets, or metrics. In the early days of symbolic AI, these benchmarks were qualitative, such as the `Blocks World' \cite{sussman1970micro,bylander1994computational}. But modern AI grounded in statistical theory has invited more sophisticated benchmark tasks, from board games like Chess \cite{campbell2002deep} and Checkers \cite{schaeffer2007checkers}, to card games like Poker \cite{bowling2015heads}, to computer games like Atari \cite{mnih2015human}, to artificial markets for testing trading algorithms \cite{wellman2001designing}, to Robocup Soccer \cite{kitano1997robocup}. 

In problems spanning across application domains such as computer vision and natural language processing (two of the most prominent application areas of modern AI techniques), the benchmark for performance on tasks has been set by widely used, standardized, large-scale evaluation datasets. For instance, in the task of object recognition, the ImageNet~\cite{russakovsky2015imagenet} dataset has become a ubiquitous standard for performance. Similarly, for the tasks of image captioning and scene understanding, the Microsoft Common Objects in Context (MS-COCO)~\cite{lin2014microsoft} dataset has been instrumental in providing a reliable benchmark for performance.

The development of such large-scale benchmark problems has led to the construction and widespread adoption of metrics to assess the performance of new algorithms at scale. Enhancements of traditional signal processing metrics such as the receiver operating characteristic (ROC)~\cite{davis2006relationship} including precision, recall, and F1 measures have been established as standard metrics for performance assessment. In more involved tasks, metrics such as mean Average Precision~\cite{van2010evaluating}(for object detection), BLEU~\cite{papineni2002bleu} (for machine translation), and Inception Score~\cite{zhou2017inception} (for generative modeling assessment) are some examples that have been established by the research community to evaluate complex intelligence tasks.



\subsection{AI as Science}



\begin{table*}[h]
\centering
\begin{tabular}{lccccc} \toprule
   \textbf{Study} & \textbf{Stimulus} & \textbf{Treatment Groups} & \textbf{AI System} & \textbf{Scope} & \textbf{\shortstack{Behavior}}\\ \midrule
 Sweeney et al., 2013 & Names as Search Terms & \shortstack{Racial Association}& \shortstack{Google Search \\Engine} & Via API & \shortstack{Disparate \\Treatment}\\
\addlinespace[1.25em]

Buolamwini et al., 2018 & \shortstack{Parliamentarian \\Headshots} & \shortstack{Gender, Fitzpatrick  \\Skin Type Class} &\shortstack{Facial Recognition \\Algorithms} & Via API & \shortstack{Disparate \\Mistreatment} \\
\addlinespace[1.25em]

Hannak et al., 2014 & Consumer Profiles &  \shortstack{ Web Broswer, Operating \\System, User History}& \shortstack{Online Pricing \\Algorithms} & \shortstack{Field} & \shortstack{Disparate \\Treatment}\\
 \bottomrule

\end{tabular}
\\[9pt]
\caption{Sample of existing behavioral studies of AI in terms of their stimulus, AI system, and measured behavior. We describe the treatment types for each stimuli and the level at which the study occurs (either locally, via an API, or ``in the field").}
\label{studies}
\end{table*}

In addition to these mathematical and engineering-based methodologies, AI researchers also frequently employ the scientific method -- specifically the hypothesis testing paradigm. For instance, to evaluate the performance of reinforcement learning algorithms in a multi-agent setting \cite{littman1994markov}, it is common to resort to Monte Carlo simulation to estimate average properties such as convergence rate and long-run performance. Likewise, in many complex AI systems, the performance of the system under different choices of parameters is often compared using the null hypothesis testing paradigm\cite{cohen1995empirical}.

However, while some AI researchers have indeed employed a hypothesis driven approach in studying AI systems, the predominant objective has been engineering-oriented: focused on designing and building intelligent systems. These pertain, for example, to the performance of a heuristic planning system against some theoretical optimum, the speed of convergence of a real-time optimization algorithm, or the classification accuracy of a given image classification system.

Indeed, it is common for articles submitted to AI conferences to be rejected if they do not present a specific technical advance beyond existing AI algorithms and models. This observation has an important implication. Even when AI researchers use the scientific method to examine the systems they build, the engineering-oriented incentives of the larger community lead them to selectively target engineering-oriented research questions. 
%
%

%
%

%
%

While these are important scientific questions about AI systems, they are not the only important questions. As we will discuss in the next section, there is a growing set of research questions of interest to a broad range of scientists outside of the scope of the AI canon that involve concepts and methods beyond the training of a typical AI researcher.


\section{A New Science of AI}
\label{newScience}
Recent advances in hardware and deep learning have led to the proliferation of deployed AI systems.  AI is no longer confined to the laboratory but instead has become a ubiquitous part of the social world. We rely on AI systems to help us make decisions as simple as what movie to watch next or where to go to dinner as well as more complex, high-stakes decisions, such as who is able to get a loan  or when an inmate’s sentence should be reduced~\cite{netflix,yelp,loan,parole}. We now share the roads with autonomous vehicles, we routinely interact with social media bots, and financial markets are now dominated by algorithmic trading \cite{cars,socialbots}. 

Due to their ubiquity and potential harm, the increasingly emergent and often unintended properties of these AI systems have garnered widespread attention in both the public and academic spheres~\cite{o2017weapons,algbias}. However, practices such as training on proprietary data and using complex models often make it challenging to use the underlying structure of the system to predict or study its emergent behaviors \cite{Voosen22}. Instead many studies, especially in the area of algorithmic bias, have employed techniques which do not require details about the system's architecture. 

In 2013, Sweeney's study on discrimination showed that search results were disproportionately more likely to return results related to arrest when a racially associated name was used as a keyword in online advertisements~\cite{sweeney2013discrimination}. This study and others like it began a new wave of hypothesis-driven science related to AI systems.

Since then, an increasing number of studies has attempted to characterize the emergent behaviors of high-stakes algorithms. A recent study showed that dark, female faces are misgendered at higher rates than lighter, male faces by commercial facial recognition algorithms \cite{buolamwini2018gender}. ProPublica showed evidence of racial discrimination in new recidivism risk score algorithms as well as price discrimination based on zip code for auto insurance premiums \cite{larson2016we,larson2017we}. Studies have even explored algorithmic bias on deployed platforms. For example, recent studies have investigated price discrimination on e-commerce sites \cite{chen2016amazon,hannak2014ecommerce}.

Engineering tools such as multi-agent simulation and ROC curves are not capable of fully explaining these systems and their behavior. Instead, we turn to the domain specific knowledge, tools, and methodologies of the social and behavioral sciences. 

%
%
%
%

%
%

\subsection{Towards a social science methodology}
%
%

\begin{figure}[t]
\centering
\includegraphics[width=0.3\textwidth]{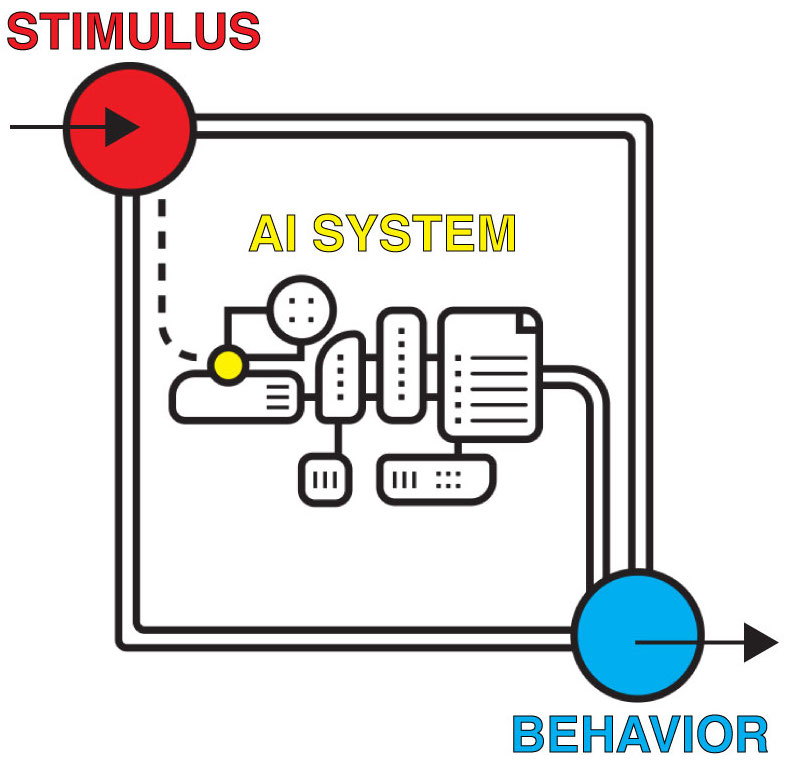}
\caption{Conceptual illustration of applying the scientific method of experimentation and causal inference to a black box AI system (center) by providing a stimulus (left) and measuring behavioral output (right).}
\label{fig:science}
\end{figure}

We propose framing the output of AI systems as behavior, with its own patterns and ecologies, and propose using scientific techniques like experimentation and causal inference to understand these behaviors, agnostic to the underlying system architecture. Figure~\ref{fig:science}  illustrates this framing. Each of the aforementioned studies in Section~\ref{newScience} contain three core components: an AI system contained in a controlled environment, systematic stimuli, and a measure of behavior, as shown in Table~\ref{studies}. 

At the core of the social sciences is the attempt to understand and predict the behavior and emergent properties of complex systems comprised of intelligent agents. Social scientists could contribute several essential pieces of expertise to questions that lie at the interface of AI and the social sciences. First, social scientists possess knowledge on experimental design and causal inference. Designing stimuli for AI systems is analogous to designing and administering more traditional `treatments' in the social and behavioral sciences. Further, in many cases such as investigating pricing discrimination on e-commerce sites, the space of potential variables is infinite. Social scientists conduct randomized controlled trials and derive theoretically informed hypotheses to narrow the space of their possible behavioral investigations. Second, social scientists bring domain specific knowledge that a typical AI researcher may not have, such as knowledge of specific types of discrimination, human learning procedures, social dynamics, among other relevant topics.

With the potential upsides to added collaboration with social and behavioral scientists in mind, we propose a new science of AI which composites previous scientific AI methodologies, the domain expertise of the social sciences, and social science methodologies such as causal inference and hypothesis-driven research design. 
%
%
%
%

\subsection{The AI Knowledge Gap}
But there are significant challenges facing this new science of AI. These challenges primarily derive from incentives which encourage the engineering of new systems over the study of existing systems. 

The first challenge relates to the reproducibility crisis,  a popular topic of discussion within the AI community. The conversation about reproducibility centers around the following problem: as researchers often desire to publish regularly and rapidly, time constraints diminish their incentives to reproduce results or replicate existing systems. The focus on contributing novel systems can  lead researchers to fail to replicate their comparative benchmarks, to choose strategic but non-comprehensive benchmarks, or to never publish the code for their novel systems. Indeed it was recently found that only 6\% of 400 authors at two top AI conferences shared their new algorithm’s code \cite{hutson2018artificial}. 

The second challenge is the expertise required to study AI systems in their social contexts. Many of the aforementioned examples require knowledge in computer science as well as knowledge (and data) related to the social domain of AI systems, such as the criminal justice system.  However, for the same reasons as with the challenge of reproducibility, there are few incentives for the AI community to gain expertise \emph{de-novo} in every domain of interest, especially for increasingly complex socio-technical systems. However, as mentioned previously, social and behavioral scientists -- due to their scientific focus on the socio-technical systems themselves -- are uniquely positioned to help fill this gap. 
%

The third challenge is accessibility.  The pervasive use of proprietary data and differences in computational power can make it difficult for computer scientists to access relevant models.  Additionally, while social scientists have unique training in measuring behavior and social outcomes, they do not necessarily have the training required to install an experimental AI system from Github or to invoke a corporate API directly. 

\begin{figure}
\includegraphics[width=0.48\textwidth]{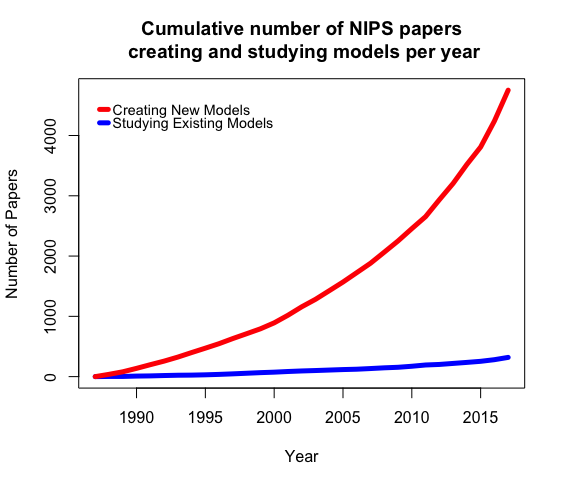}
\caption{Proxy evidence for an \emph{AI Knowledge Gap}: the number of AI agents built grows faster than the studies that characterize those agents' behavior.}
\label{fig:divergence}
\end{figure}

These dynamics have caused the study of existing AI systems to severely lag behind the production of new systems, as shown in Figure~\ref{fig:divergence}. To estimate the pace of algorithm development in machine learning and artificial intelligence, we study the full set of 7,241 conference papers from the Neural Information Processing Systems (NIPS) Conference from 1987 to 2017. To count the number of papers introducing new computational models, we search through abstracts and count the number of papers that contain any of the words ‘new’, ‘novel’, ‘we propose’, and ‘we introduce’ in their abstract. To count the papers aimed solely at studying previous models, we count the number of papers that include any of the terms ‘we study’, ‘we examine’, ‘we compare’ and ‘we analyze’ in their abstract, but none of the keywords used to count new models. 

As discussed above, the key bottleneck in the study of the behavior of AI systems is a misalignment of incentives. To address these concerns, we propose TuringBox, a market platform that provides access to and incentive for the study of AI behavior. 

\section{TuringBox: A Market for the Behavioral Research of AI Systems}

In order to address the above challenges of accessibility, replication, and incentive incompatibility, a general tool for the rigorous study of AI systems must satisfy many properties. Drawing from the literature of existing platforms for AI research, we enumerate these desirable properties below:
\begin{enumerate}
\item \textbf{Customizable Metrics:} The platform should allow for the creation of custom evaluation metrics to evaluate AI systems, and the customizable selection of subsets of stimuli (data samples or environments). For instance, users might wish to evaluate custom metrics of fairness, precision, and accuracy on a variety of algorithms for a particular task, or to evaluate algorithms on a variety of subsets of stimuli. 
\item \textbf{Centralized Implementation:} Scientists on the platform should be able to evaluate the exact implementation on any arbitrary input sample, along with explicitly outlined hyperparameter settings that produced the model. This property has been adopted by platforms such as ParlAI~\cite{miller2017parlai} and Algorithmia.
\item \textbf{Centralized Evaluation:} 
The platform should unify several different benchmarks that exist across communities in a centralized computing environment, without replicating the benchmarks locally. Such an evaluation system has already been adopted by platforms such as ParlAI~\cite{miller2017parlai}, Algorithmia and OpenML~\cite{vanschoren2014openml}. 
\item \textbf{General Compatibility:} The platform should have a consistent, generalizable method to include newer AI problems and have support for all existing AI problems of interest. This property is essential to understand the behavior of classes of algorithms across different tasks and environments, and provides a cohesive framework for the study of AI behavior across different classes of algorithms. 
\item \textbf{Full Scope:} AI systems of interest can be accessed at three different scopes. The first scope consists of algorithms whose code is readily available and can be accessed directly from a scientist's local machine or from a centralized platform's servers. The second scope includes algorithms that are accessible only from remote servers via API calls. The third scope contains remotely served algorithms that do not have an explicit API protocol, and thus must be observed ``in the field," via custom scripts, as exemplified in Hannak et al. A sufficient platform for the behavioral study of AI should provide functionality for each scope. 
\end{enumerate}

\begin{figure}[t]
\includegraphics[width=0.5\textwidth]{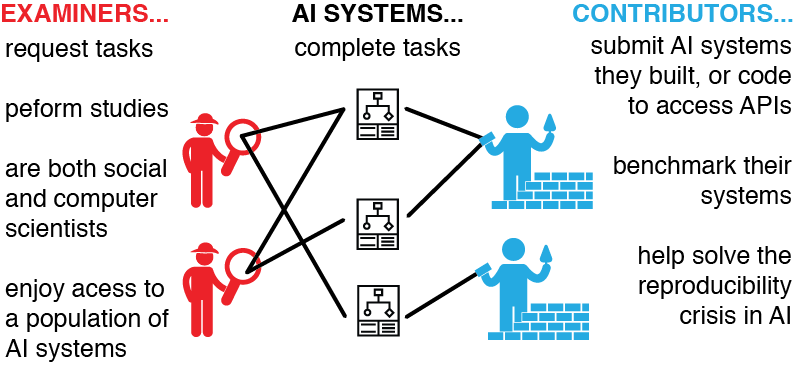}
\caption{A schematic of the marketplace for AI.}
\label{market}
\end{figure}

With these properties in mind, we envision a two-sided digital marketplace, called TuringBox, that couples two previously orthogonal threads of AI research by convening two communities (see Fig~\ref{market}). On one side of the market, \textit{AI contributors} upload existing and novel algorithms to both benchmark their algorithms with respect to performance, fairness, or other qualities and to gain reputation in their community. These contributors will be incentivized to upload not only algorithms they wrote themselves, but also protocols that access APIs and algorithms ``in the field.''

On the other side of the market, \textit{AI examiners} develop and post tasks designed to examine and characterize algorithmic behavior. We anticipate that this methodological tool will attract not just computer scientists but will also be of interest to experts in experimentation across the social and behavioral sciences, thus mitigating the AI Knowledge Gap. 

\begin{table*}[h]
\centering
\begin{tabular}{lccccc} \toprule
 Platform & \shortstack{Customizable \\ Metrics}  & 
 \shortstack{Centralized \\Implementation}
 & \shortstack{Centralized \\Evaluation}& \shortstack{General \\Compatibility} &  \shortstack{Full \\Scope} \\ [0.5ex] 
\midrule
\addlinespace[0.5em]
  Algorithmia & \checkmark &  \checkmark &  \checkmark &  \checkmark& \\ \hline
  \addlinespace[0.5em]

  CloudCV &  & &  \checkmark &&   \\ \hline
  \addlinespace[0.5em]

 DeepMind PsychLab & \checkmark & & & &\\ \hline
 \addlinespace[0.5em]

 FairML &   & &  &  \checkmark & \\ \hline
 \addlinespace[0.5em]

 OpenML &  \checkmark &  & \checkmark & \checkmark&\\ \hline
 \addlinespace[0.5em]

 ParlAI &  \checkmark &  \checkmark &  \checkmark & & \\ \hline
 \addlinespace[0.5em]

 Themis-ML &  &  \checkmark &  \checkmark &  \checkmark& \\ \hline
 \addlinespace[0.5em]

\textbf{TuringBox} &  \checkmark &  \checkmark &  \checkmark &  \checkmark &\checkmark\\ 

 \bottomrule
\end{tabular}
\caption{Platforms for the study of AI and their corresponding properties.}
\end{table*}

Notable examples of AI platforms include the DeepMind PyschLab~\cite{leibo2018psychlab}, a fully 3D game-like platform for agent-based AI research, ParlAI~\cite{miller2017parlai}, a unified framework for the evaluation of dialog models with additional support for images, CloudCV~\cite{agrawal2015cloudcv}, an open-source platform for machine learning and computer vision, and OpenML~\cite{vanschoren2014openml}, an online machine learning platform for sharing and organizing data and machine learning algorithms. Each platform satisfies some of the properties listed above, and we map these platforms to properties in Table 2. 


\subsection{Incentive Scheme}

In our proposed market, both sides are driven by exogenous and endogenous incentives rather than money. The implementation is engineered to best serve the unique interests of both types of users.  Below we discuss how each side of the market incentivizes use by the other. 


\subsubsection{Incentives of examiners}


As AI systems further permeate society, we envision increasing demand for the examination of the behaviors of these systems. The recent increase in studies on algorithmic bias~\cite{hajian2016algorithmic} as well as the success of interdisciplinary conferences such as Fairness, Accountability, and Transparency or the International Conference on Computational Social Science already suggest an interest from the social science community. In addition to understanding bias, examiners may also be able to learn about the origins and structure of complex human behavior, such as strategy formation, as was the case with AlphaGo Zero~\cite{silver2016mastering}. 

However, accessibility is a key concern preventing examiners from studying existing systems. Many social scientists have limited programming or computer science backgrounds and cannot implement state of the art AI systems to study. To address this concern, we employ the second side of the market (the contributors) who upload their AI systems to the platform. Examiners can then access these systems from a central source without added expertise. 
%
%

\subsubsection{Incentives of the contributors}

TuringBox motivates a large group of people -- the contributors -- to populate the platform with algorithms to study. These contributors are AI researchers in academia and industry as well as other individuals that possess the skills to design and implement AI systems. 

TuringBox will motivate contributors to upload their own systems by addressing several concerns in the AI community. Importantly, the platform will assist with the problem of reproducibility.  When contributors upload their new systems, all other contributors will then be able to access and test these new systems. Having the designers of the algorithms themselves upload their algorithm is important for two reasons. First, it prevents issues from arising due to differences in implementations. Secondly, it ensures that the algorithm designers have responsibility for the performance of the algorithm and any subsequent aberrant behavior it may exhibit.

Further, TuringBox will ease access to benchmarking tools. With so many AI systems on the platform, contributors will be able to easily compare their system's performance to the performance of other state of the art systems. Moreover,  by incorporating social scientists on the examiner side of the market, TuringBox will allow researchers to measure their systems using new social metrics, such as fairness. As examiners study an increasing number of AI systems, we anticipate they will produce new ways to measure the biases or behaviors of AI systems, which contributors can then use to ensure their systems operate as expected.



In addition to individual contributors, companies  may also be interested in  filling the contributor side of the market via APIs to demonstrate the performance of their proprietary algorithms and that their algorithms are, for example, bias-free.

\section{Discussion}
Because of AI systems' emergent complexity, their ubiquity in society, and their inherent opacity,  there is a need to map the boundaries of AI research, and to extend them when possible. Our examination suggests there is untapped potential for the hypothesis driven scientific investigation of AI systems. It is useful to think of  AI systems not solely as engineering artifacts, but rather as a new class of protocols with heterogeneous behavior. In order to keep up with the proliferation of these systems and to close the AI Knowledge Gap, AI scientists must start to emphasize the generation of knowledge about how these systems behave both in the lab and in the field. This can be achieved by broadening the scope of AI research and opening access to the study of AI behavior to scientists from other disciplines. 

We believe a market for the study of machine behavior is crucial to reach these goals for several reasons. The first is that it offers a general yet unified framework for understanding when bias occurs in complex architectures. Many studies across computer science, behavioral economics, and legal studies have already investigated the behavior of AI systems but each uses an ad hoc approach to collecting data and measuring behavior. Our market provides a toolkit for examining  behavior across a population of AI systems, which enables a scalable and flexible alternative to costly algorithmic audits.  As an increasing number of AI systems are deployed every day, and their real world stakes increase, a consistent methodology to perform these algorithmic audits at scale becomes imperative. 

The second rationale for the TuringBox platform is that it helps bridge the gap between computer scientists and social scientists. Historically, computer scientists and roboticists were the only ones concerned with the behavior of machines, because they were the only ones interacting with the systems. However, social sciences offer many key skills, methods, and perspectives about AI systems that are not being fully leveraged in the AI research community. 

Finally, this framing can potentially curb adversarial effects of emergent superintelligences by providing a controlled environment to examine their behaviors \cite{bostrom1998long}. Indeed this market anticipates a future where the AI systems that inhabit it exhibit the complex, cross-domain behaviors of artificial general intelligence. 

Amazon’s Mechanical Turk created a revolution for the social sciences by scaling the way social scientists  perform experiments~\cite{buhrmester2011amazon,horton2011online}. For the first time, scientists could cheaply run massive online experiments and learn about individual and collective behavior without recruiting subjects to a physical location. By lowering the barrier to entry for experimentation, the market was successful in both opening up new methodologies and democratizing research to a broad class of scientists. As a behavioral science of AI systems becomes conceivable, we believe the next revolution in AI experimentation will come from standardizing experimental protocols  while keeping the heterogeneity of these systems wide open. 
\appendix

\bibliographystyle{named}
\bibliography{ijcai18}

\end{document}